\documentclass[12pt, leqno]{article}
\usepackage{amsmath,amsfonts, amssymb, color,hyperref}

\numberwithin{equation}{section}

\newtheorem{theorem}{Theorem}[section]
\newtheorem{definition}[theorem]{Definition}
\newtheorem{proposition}[theorem]{Proposition}
\newtheorem{corollary}[theorem]{Corollary}
\newtheorem{lemma}[theorem]{Lemma}

\newcommand{\ang}[1]{\langle  #1 \rangle }  

\newcommand{\p}{\mathbb{P}}
\newcommand{\Q}{\mathbb{Q}}

\newcommand{\E}[1]{\mathbb{E}\left[#1\right]}
\newcommand{\Et}[1]{\mathbb{E}_t\left[#1\right]}

\newcommand{\trieq}{\stackrel{\triangle}{=}}
\newcommand{\eof}{\hfill {\it Q.E.D.} \vspace*{0.3cm}}

\newcommand{\pf}{{\it Proof: }}

\newcommand{\cF}{{\mathcal F}}
\newcommand{\cG}{{\mathcal G}}

\newcommand{\R}{\mathbb{R}}
\newcommand{\id}{{\mathbf 1}}

\title{{Time-Inconsistent Stochastic Linear--Quadratic Control: Characterization and Uniqueness of Equilibrium}}

\author{Ying Hu\thanks{IRMAR,
Universit\'e Rennes 1, 35042 Rennes Cedex, France. The research of this author was  partially supported by Lebesgue center of mathematics ``Investissements d'avenir" program - ANR-11-LABX-0020-01.} \and
Hanqing Jin\thanks{Mathematical Institute and Nomura Centre for Mathematical Finance, and
Oxford--Man Institute of Quantitative Finance, The University of
Oxford, Oxford OX2 6GG, UK. The research of this author was  partially supported by  research
grants from the Nomura Centre for
Mathematical Finance and the Oxford--Man Institute of Quantitative Finance. }
\and
Xun Yu Zhou\thanks{Mathematical Institute and Nomura Centre for Mathematical Finance, and
Oxford--Man Institute of Quantitative Finance, The University of
Oxford,  Oxford OX2 6GG, UK. The research of this author was supported by  a start-up
fund of the University of Oxford, and research grants from the Nomura Centre for
Mathematical Finance and the Oxford--Man Institute of Quantitative Finance.}}

\date{April 28, 2015}
\begin{document}
\maketitle

\begin{abstract}
In this paper, we continue our study on a general time-inconsistent stochastic linear--quadratic
(LQ) control problem originally formulated in \cite{HJZ}. We derive a necessary and sufficient condition
for equilibrium controls via a flow of forward--backward stochastic differential equations. When the
state is one dimensional and the coefficients in the problem are all deterministic, we
prove that the explicit
equilibrium control constructed in \cite{HJZ} is indeed unique. Our proof is based on the derived equivalent condition for equilibria as well as a stochastic version of the Lebesgue
differentiation theorem. Finally,  we show that the  equilibrium strategy is unique for a mean--variance portfolio selection model
in a complete financial market where the risk-free rate is a deterministic function of time but all the
other market parameters are possibly stochastic processes.

\end{abstract}


{\bf AMS subject.}
 91B51, 93E99, 60H10

\section{Introduction}
Time inconsistency in dynamic decision making is often observed in social systems and daily life. The study
on time inconsistency  by economists dates back to  Strotz \cite{Strotz} in the 1950s, who proposed the formulation of  a time-inconsistent decision problem as a game between incarnations  of the controller at different time instants.

The game formulation is fairly straightforward and easy to understand when the time setting is discrete. In a continuous-time setup, the formulation can be generalized in different ways.
Yong \cite{Yong} and
 Ekeland and Pirvu \cite{EP}  define equilibrium controls in the class of feedback
 policies for problems involving hyperbolic discounting, and prove the existence of equilibria.
 Grenadier and Wang \cite{GW} investigate optimal stopping with, again,  hyperbolic discounting.
Bj\"ork and Murgoci \cite{BM} formulate a general Markovian stochastic control problem with time inconsistent terms, and establish
sufficient conditions for equilibria through a generalized HJB equation system. They then present some special cases including a linear--quadratic (LQ) control  problem in which solutions are constructed. Bj\"ork, Murgoci and Zhou \cite{BMZ} further derive analytically an equilibrium strategy for a mean--variance portfolio selection model with state-dependent risk aversion.

In our previous paper, \cite{HJZ}, we formulate a general non-Markovian stochastic LQ control problem, where the
objective functional includes terms leading to time-inconsistency, and derive a general
{\it sufficient} condition for equilibria through a system of forward--backward stochastic
differential equations (FBSDEs). Based on this condition, we construct explicitly an equilibrium control when the state is scalar-valued and all the coefficients are non-random.
In contrast to the aforementioned works where an equilibrium control is defined within the class of feedback
controls, we define our equilibrium via open-loop controls.

Most of the existing literature on game formulation of time-inconsistent problems
has focused on existence of equilibria, and the only paper according to our best knowledge
that mentions about the uniqueness is Vieille and Weibull \cite{VW}, in which the authors   show that the uniqueness does {\it not} hold in a discrete-time model.
Uniqueness is important in both practice and theory. In applications, multiple equilibria lead to multiple value processes,\footnote{The value of an equilibrium control is the corresponding objective
 functional value.} and there is an issue of the choice of the one to use and implement.
Theoretically speaking, when a new, {\it weak} notion of a solution is introduced the uniqueness is always important, for it is one of the touchstones of the appropriateness of the new definition (non-uniqueness is a sign that the notion may be {\it too weak} to be useful). On the other hand, mathematically, proving uniqueness of a weaker notion is almost always
challenging.\footnote{A good example is the uniqueness of viscosity solution for a nonlinear PDE; see \cite{YZ}.}

In this paper, we take on the challenge of establishing the uniqueness of equilibrium control for the same time-inconsistent model formulated in \cite{HJZ}. First, we  derive a general {\it necessary and sufficient} condition
for equilibrium controls. A key step in the derivation is to prove a stochastic version of the Lebesgue
differentiation theorem which is interesting in its own right and potentially useful
for other stochastic control problems. Then, we focus on the case in which the
state is one dimensional and the coefficients in the problem are all deterministic.
Thanks to the derived equivalent condition for equilibria we
show that the explicit
equilibrium control constructed in \cite{HJZ} is indeed unique. Finally,  we prove that the  equilibrium strategy, again constructed in \cite{HJZ},  is unique for a mean--variance portfolio selection model
in a complete financial market where the risk-free rate is a deterministic function of time but all the
other market parameters are possibly stochastic processes.


The rest of this paper is organized as follows. In Section \ref{prob-formu}, we recall the formulation of the
time-inconsistent LQ control problem  studied in our previous work \cite{HJZ}.
We then derive  an equivalent characterization of equilibrium controls in terms of
 the solution to a system of FBSDEs in Section \ref{formal-derivation}. In Section \ref{sec-uniq}
 we prove that the equilibrium obtained in \cite{HJZ} is the unique one. Section \ref{mvcase} is devoted to the uniqueness for a mean--variance portfolio selection model. Finally, Section \ref{concl}
 concludes. Some technical derivations are placed in appendices.

\section{Problem Formulation}\label{prob-formu}

Let $(W_t)_{0 \le t \le T}=(W_t^1,\cdots,W_t^d)_{0 \le t \le T}$
be a $d$-dimensional Brownian motion on a probability
space $(\Omega, \cF, \p)$. Denote by $(\cF_t)$ the
augmented filtration generated by $(W_t)$.

We will use the same notation as in our previous paper \cite{HJZ}, which we list here for the reader's convenience:
\begin{table}[http]
\begin{tabular}{rl}
 $\mathbb S^l$:& the set of symmetric $l \times l$ real matrices.\\
  $L^2_{\cG}(\Omega; \, \R^l)$:& the set of random variables $\xi: (\Omega, \cG) \rightarrow (\R^l, {\cal B}(\R^l))$ \\ &with $\E{|\xi|^2}<+\infty$.\\
 $L^\infty_{\cG}(\Omega; \, \R^l)$:& the set of essentially bounded  random variables \\  &$\xi: (\Omega, {\cG}) \rightarrow (\R^l, {\cal B}(\R^l))$.\\
 $L^2_\cG(t, \, T; \, \R^l)$: &the set of $\{\cG_s\}_{s\in [t,T]}$-adapted processes \\ &$f=\{f_s: t\leq s\leq T\}$ with $\E{ \int_t^T|f_s|^2\, ds} < \infty$.\\
 $L^\infty_\cG(t, \, T; \, \R^l)$:&the set of essentially  bounded $\{\cG_s\}_{s\in [t,T]}$-adapted processes.\\
 $L^2_\cG(\Omega; \, C(t, \, T; \, \R^l))$:& the set of continuous
         $\{\cG_t\}_{s\in [t,T]}$-adapted processes \\&$f=\{f_s: t\leq s\leq T\}$ with $\E{ \sup_{s\in [t,T]}|f_s|^2\, } < \infty$.
\end{tabular}
\end{table}

We will often use vectors and matrices in this paper, where all vectors are column vectors. For a matrix $M$,
 define
 \begin{table}[http]
\begin{tabular}{rl}
$M'$: &transpose of a matrix $M$.\\
 $|M|=\sqrt{\sum_{i,j}m_{ij}^2}$: &Frobenius norm of a matrix $M$.
\end{tabular}
\end{table}

The time-inconsistent LQ control model under consideration in this paper was introduced
in \cite{HJZ}. Here we recall the formulation.

Let $T>0$ be given and fixed. The controlled system is governed by the following stochastic
differential equation (SDE)
on $[0, T]$:
\begin{equation}\label{controlgeneral:t}
dX_s=[A_sX_s+B_s'u_s+b_s]ds+ \sum_{j=1}^d[C_s^jX_s+D_s^{j} u_s+\sigma_s^{j} ]dW_s^{j};\quad X_0=x_0,
\end{equation}
where  $A$ is a bounded deterministic  function on $[0, T]$ with value in $\R^{n\times n}$,
$B,C^j,D^j$ are all essentially bounded adapted processes on $[0,T]$
with values in $\R^{ l\times n}$,
$\R^{n\times n}$, $\R^{ n\times l}$,  respectively, and  $b$ and $\sigma^j$ are stochastic processes
in  $L^2_\cF(0,T; \R^n)$.
The process $u\in L^2_\cF(0, \, T; \, \R^l)$ is the control, and
$X\in L^2_\cF(\Omega; \, C(0, \, T; \, \R^n))$ is the corresponding state process  with initial value $x_0\in \R^n$.

When time evolves to $t\in[0,T]$, we need to  consider the controlled system
starting from  $t$ and state $x_t\in L^2_{\cF_t}(\Omega; \, \R^n)$:
\begin{equation}\label{controlgeneral:t}
dX_s=[A_sX_s+B_s'u_s+b_s]ds+ \sum_{j=1}^d[C_s^jX_s+D_s^{j} u_s+\sigma_s^{j} ]dW_s^{j}, \quad X_t=x_t.
\end{equation}
For any  control $u\in L^2_\cF(t, T;  \R^l)$, there exists
a unique  solution $X^{t,x_t,u}\in L^2_\cF(\Omega; \, C(t,  T;  \R^n))$.
At $t$ with the system state $X_t=x_t$, our aim  is to minimize
\begin{eqnarray}\label{costgeneral}
J(t,x_t;u)&\trieq&\frac{1}{2}\mathbb E_t\int_t^T\left[ \ang{Q_sX_s, X_s}+\ang{ R_su_s, u_s}\right]ds+\frac{1}{2}\mathbb E_t [\ang{G X_T, X_T}]\nonumber \\
&&- \frac{1}{2} \ang{h\Et{X_T}, \Et{X_T}}-\ang{ \mu_1 x_t+\mu_2,  \Et{X_T}}
\end{eqnarray}
over  $u\in L^2_\cF(t, \, T; \, \R^l)$,
where $X=X^{t,x_t,u}$, and $\Et{\cdot}=\E{\cdot|\cF_t}$.
In the above $Q$ and $R$ are both positive semi-definite and essentially bounded adapted processes on $[0,T]$ with values in ${\mathbb S}^n$
and ${\mathbb S}^l$ respectively,  $G, h, \mu_1, \mu_2$ are constants in $\mathbb S^n$, $\mathbb S^n$,   $\R^{n\times n}$, $\R^n$ respectively, and moreover $G$ is  positive semi-definite.

We define an {\it equilibrium (control)} in the following manner.
Given a  control $u^*$, for any $t\in [0,T)$, $\varepsilon>0$ and $v\in L^2_{\cF_t}(\Omega; \, \R^l)$,  define
\begin{equation}\label{svgeneral}
u^{t,\varepsilon,v}_s=u^*_s+v\id_{s\in [t,t+\varepsilon)},\;\;\;s\in[t,T].
\end{equation}

\begin{definition}
Let $u^*\in L^2_\cF(0, \, T; \, \R^l)$ be a given control and $X^*$ be the state process corresponding to $u^*$. The control $u^*$ is called an equilibrium
 if
$$\liminf_{\varepsilon\downarrow 0} \frac{J(t,X^*_t; u^{t,\varepsilon,v})-J(t,X^*_t;u^*)}{\varepsilon}\ge 0,$$
where $u^{t,\varepsilon,v}$ is defined by (\ref{svgeneral}), for any $t\in [0,T)$ and $v\in L^2_{\cF_t}(\Omega; \, \R^l)$.
\end{definition}
Notice that here we have changed $\lim$ in \cite{HJZ} to $\liminf$ in this definition, resulting in a weaker definition. As a result, the sufficient condition derived in \cite{HJZ} is also sufficient for this new definition. On the other hand, the uniqueness result to be established for the new definition will also imply the uniqueness for the old one. For these reasons, the above definition appears to be more appropriate.

\section{Necessary and Sufficient Condition  of Equilibrium Controls}\label{formal-derivation}

In our previous paper, \cite{HJZ}, a sufficient condition is derived via
the second-order expansion in the local spike variation, in the
same spirit of proving the stochastic Pontryagin's maximum principle
\cite{PBGM, peng90,YZ}. In this section, we present a general necessary and sufficient condition for equilibria. This condition is made possible by a stochastic Lebesgue differentiation theorem involving conditional expectation. The latter theorem, interesting in its own right, is new according to our best knowledge.

To proceed, we start with some relevant known result from our previous paper \cite{HJZ}.
Let $u^*$ be a fixed control and $X^*$ be the corresponding state process.
For any $t\in [0, T)$, define in the time interval $[t, T]$  the  processes
$(p(\cdot;t), (k^j(\cdot;t))_{j=1,\cdots, d})\in L^2_\cF(t,T;\R^n)\times (L^2_\cF(t,T;\R^n))^d$
as  the unique solution to
\begin{equation} \label{adjoint1general}
\left\{\begin{array}{l}
dp(s;t)=-[A_s'p(s;t)+\sum_{j=1}^d(C_s^j)' k^j(s;t)+Q_sX^*_s]ds\\
\;\;\;\;\;\;\;\;\; +\sum_{j=1}^dk^j(s;t)dW_s^j,\;\;s\in[t,T],\\
p(T;t)=G X^*_T- h \Et{X^*_T}-\mu_1 X_t^*-\mu_2.
\end{array}\right.
\end{equation}

Furthermore, define  $(P(\cdot;t), (K^j(\cdot;t))_{j=1,\cdots,d})\in L^2_\cF(t,T;\mathbb S^n)\times (L^2_\cF(t,T;\mathbb S^n))^d$ as  the unique solution to
\begin{equation}\label{adjoint2general}
\left\{\begin{array}{ll}
dP(s;t)=&-\Big\{A_s'P(s;t)+P(s;t)A_s \\
&+\sum_{j=1}^d[(C_s^j)'P(s;t)C_s^j+(C_s^j)'K^j(s;t)+K^j(s;t)C_s^j]+Q_s\Big\}ds\\
&+\sum_{j=1}^dK^j(s;t)dW_s^j,\;\;s\in[t,T],\\
P(T;t)=&G.
\end{array}\right.
\end{equation}

The following estimate under local spike variation is reproduced from \cite[Proposition 3.1]{HJZ}.
\begin{proposition}\label{variate}
 For any $t\in [0,T)$, $\varepsilon>0$ and $v\in L^2_{\cF_t}(\Omega; \, \R^l)$,  define $u^{t,\varepsilon,v}$
by (\ref{svgeneral}). Then
\begin{equation}\label{epsilongeneral}
J(t,X^*_t; u^{t,\varepsilon,v})-J(t,X^*_t; u^*)
=\mathbb E_t\int_t^{t+\varepsilon} \left(\ang{\Lambda(s;t),v}
+\frac{1}{2}\ang{H(s;t)v, v}\right)ds+o(\varepsilon)
\end{equation}
where $\Lambda(s;t)\trieq   B_sp(s;t)+\sum_{j=1}^d(D_s^j)'k^j(s;t)+R_su^*_s$ and
$H(s;t)\trieq R_s+\sum_{j=1}^d (D_s^j)'P(s;t)D_s^j$.
\end{proposition}


In view of Proposition \ref{variate} and the fact that $H(s;t)\succeq 0$, it is straightforward to get the following characterization of an equilibrium.
\begin{corollary}\label{equiv-lq}
A control $u^{*}\in L^{2}_\cF(0,T, \R^{l})$ is an equilibrium if and only if
\begin{equation}\label{cond1}
\lim_{\varepsilon\downarrow 0}\frac{1}{\varepsilon}\int_{t}^{t+\varepsilon}\Et{\Lambda(s;t)}ds=0,  \;\;a.s., \;\; \forall t\in [0,T).
\end{equation}
\end{corollary}

The next result provides a key property for the solution to the
BSDE (\ref{adjoint1general}), and represents the process
$\Lambda(s;t)$ in a special form.

\begin{proposition}\label{k-ind-t}
For any given pair of state and control processes $(X^{*}, u^{*})$, the solution to  (\ref{adjoint1general}) satisfies
$k(s;t_{1})=k(s;t_{2})$ for a.e. $s\ge \max{( t_{1},t_{2})}$.
Moreover, there exist $\lambda_{1}\in L^{2}_{\cF}(0, T; \R^{l}), \lambda_{2}\in L^{\infty}_{\cF}(0, T; \R^{l\times n}) $ and $\xi\in L^{2}(\Omega; C(0, T; \R^{n}))$, such that
$\Lambda(s;t)$ has the representation
$$\Lambda(s;t)=\lambda_{1}(s)+\lambda_{2}(s)\xi_{t}.$$
\end{proposition}
\pf Define the function $\psi(\cdot)$ as the unique solution for the matrix-valued ordinary differential equation (ODE)
$$d\psi(t)=\psi(t)A(t)'dt, \;\;\;\; \psi(T)=I_{n},$$
where $I_{n}$ denotes the $n\times n$ identity matrix.  It is clear that
$\psi(\cdot)$ is invertible, and both $\psi(\cdot)$ and $\psi(\cdot)^{-1}$ are  bounded.

Let $\hat p(s;t)=\psi(s) p(s;t)+h\Et{X^{*}_{T}}+\mu_{1}X^{*}_{t}+\mu_{2}$
 and $\hat k^{j}(s;t)=\psi(s) k^{j}(s;t)$ for $j=1,\cdots, d$.
Then  on the interval $[t, T]$,  $(\hat p(\cdot;t), \hat k(\cdot;t))$ satisfies
\begin{equation}\label{adjoint1tilde}
\left\{\begin{array}{l}
d\hat p(s;t)=-\left[\sum_{j=1}^d\psi(s)(C_s^j)' \psi(s)^{-1}\hat k^j(s;t)+\psi(s)Q_s X^*_s\right]ds+\sum_{j=1}^d\hat k^j(s;t)dW_s^j,\\
\hat p(T;t)=G X^*_T.
\end{array}\right.
\end{equation}
Notice that neither the terminal condition nor the coefficients of this equation depend on
$t$; so it can be taken as a BSDE on the entire time interval $[0,T]$. Denote its solution as
$(\hat p(s), \hat k(s))$, $s\in[0,T]$. It then follows from the
uniqueness of the solution to BSDE that $(\hat p(s;t), \hat k(s;t))=(\hat p(s), \hat k(s))$
at $s\in[t,T]$ for any $t\in[0,T]$.
As a result, $k(s;t)=\psi(s)^{-1} \hat k(s):=k(s)$, proving the first claim of
the proposition.

Next,
 $$p(s;t)=\psi(s)^{-1}\hat p(s)-\psi(s)^{-1}(h\Et{X^{*}_{T}}+\mu_{1}X^{*}_{t}+\mu_{2})
 =p(s)+\psi(s)^{-1}\xi_{t},$$
 where 
 $\xi_{t}:=-h\Et{X^{*}_{T}}-\mu_{1}X^{*}_{t}-\mu_{2}$ defines the process $\xi\in L^{2}_{\cF}(\Omega; C(0, T; \R^{n}))$ and $p(s):=\psi(s)^{-1}\hat p(s)$ defines the process $p\in L^{2}_{\cF}(\Omega; C(0, T; \R^{n}))$. Consequently,
 \begin{eqnarray*}
 \Lambda(s;t)&=&B_sp(s;t)+\sum_{j=1}^d(D_s^j)'k^j(s;t)+R_su^*_s\\
 &=&B_sp(s)+\sum_{j=1}^d(D_s^j)'k^j(s)+R_su^*_s+B_{s}\psi(s)^{-1}\xi_{t}\\
 &=&\lambda_{1}(s)+\lambda_{2}(s)\xi_{t},
\end{eqnarray*}
where $\lambda_{1}(s):=B_sp(s)+\sum_{j=1}^d(D_s^j)'k^j(s)+R_su^*_s$ and $\lambda_{2}(s):=B_{s}\psi(s)^{-1}$.
 \eof

We now set out to derive our general necessary and sufficient condition for equilibrium
controls. Although (\ref{cond1}) already provides a characterizing condition, it is nevertheless not very useful
because it involves a limit. It is tempting to expect that the limit therein is $\Lambda(t;t)$,
in the spirit of the Lebesgue differentiation theorem.\footnote{A simple version of this theorem states that
if $\varphi$ is an integrable real function on $[0,T]$, then $\lim_{\varepsilon\downarrow 0}\frac{1}{\varepsilon}\int_{t}^{t+\varepsilon}\varphi(s)ds=\varphi(t)$ a.e. $t\in [0,T]$.} However,
one needs to be very careful since in (\ref{cond1}) the conditional expectation with respect to ${\cal F}_t$ is
involved. 
The following general result can be regarded as a {\it stochastic} Lebesgue differentiation theorem. While it serves our purpose in this paper, it is of interest in
its own right and may be potentially useful for (among others) various stochastic control problems.

\begin{lemma}\label{cond-lebesgue}
Let $Y\in L^{2}_{\cF}(0,T; \R^l)$ be a given process. If $\lim_{\varepsilon\downarrow 0}\frac{1}{\varepsilon}\int_{t}^{t+\varepsilon}\Et{Y_s}ds=0,\;  a.e. t\in [0, T), a.s.$,
then $Y_t=0,\;  a.e. t\in [0, T), a.s.$.
\end{lemma}
\pf  Since $L^{2}_{\cF_{T}}(\Omega; \R^l)$ is a separable space, it follows from
the (deterministic) Lebesgue differentiation theorem that there is a countable dense subset  ${\mathcal D}\subset  L^{2}_{\cF_{T}}(\Omega;  \R^l)\cap L^{\infty}_{\cF_{T}}(\Omega;  \R^l)$,
such that for almost all $t$, we have
\begin{equation}\label{limtoprod}
\lim_{\varepsilon\downarrow 0}\frac{1}{\varepsilon}\int_{t}^{t+\varepsilon}\E{\langle Y_{s},\eta\rangle }ds=\E{\langle Y_{t},\eta\rangle }, \quad \forall \eta\in {\mathcal D},
\end{equation}
and $\lim_{\varepsilon\downarrow 0}\frac{1}{\varepsilon}\int_{t}^{t+\varepsilon}\E{Y_{s}^{2}}ds=\E{Y_{t}^{2}}.$

For any $\eta\in {\mathcal D}$, define $\eta_{s}={\mathbb E}_{s}[\eta]$. Then
$\E{\langle Y_{s},\eta\rangle }=\E{\langle Y_{s},\eta_s\rangle }$. We have the following estimates:
\begin{eqnarray*}
\left|\lim_{\varepsilon\downarrow 0}\frac{1}{\varepsilon}\int_{t}^{t+\varepsilon}\E{\langle Y_{s}, \eta_{s}-\eta_{t}\rangle }ds\right|
&\le&\lim_{\varepsilon\downarrow 0}\frac{1}{\varepsilon}\sqrt{ \int_{t}^{t+\varepsilon}\E{Y_{s}^{2}}ds \int_{t}^{t+\varepsilon}\E{(\eta_{s}-\eta_{t})^{2}}ds }\\
&=&\lim_{\varepsilon\downarrow 0}\sqrt{\frac{1}{\varepsilon} \int_{t}^{t+\varepsilon}\E{Y_{s}^{2}}ds}\sqrt{ \frac{1}{\varepsilon}\int_{t}^{t+\varepsilon}\E{(\eta_{s}-\eta_{t})^{2}}ds }\\
&\le&\lim_{\varepsilon\downarrow 0}\sqrt{\frac{1}{\varepsilon} \int_{t}^{t+\varepsilon}\E{Y_{s}^{2}}ds }\sqrt{\sup_{s\in [t,t+\varepsilon]}\E{(\eta_{s}-\eta_{t})^{2}} }\\
&\le&2\lim_{\varepsilon\downarrow 0}\sqrt{\frac{1}{\varepsilon} \int_{t}^{t+\varepsilon}\E{Y_{s}^{2}}ds} \sqrt{\E{(\eta_{t+\varepsilon}-\eta_{t})^{2}} }
=0,
\end{eqnarray*}
where the last inequality is due to Doob's martingale inequality as $\eta_s$ is a square-integrable martingale.
Hence for any $\eta\in {\mathcal D}$,
\begin{eqnarray*}
\E{\langle Y_{t},\eta_{t}\rangle }&=&\E{\langle Y_{t},\eta\rangle }\\
&=&\lim_{\varepsilon\downarrow 0}\frac{1}{\varepsilon}\int_{t}^{t+\varepsilon}\E{\langle Y_{s}, \eta\rangle }ds\\
&=&\lim_{\varepsilon\downarrow 0}\frac{1}{\varepsilon}\int_{t}^{t+\varepsilon}\E{\langle Y_{s}, \eta_{s}\rangle }ds\\
&=&\lim_{\varepsilon\downarrow 0}\frac{1}{\varepsilon}\int_{t}^{t+\varepsilon}\E{\langle Y_{s}, \eta_{t}\rangle }ds\\
&=&\lim_{\varepsilon\downarrow 0}\frac{1}{\varepsilon}\int_{t}^{t+\varepsilon}\E{\langle \Et{Y_{s}}, \eta_{t}\rangle }ds\\
&=&\lim_{\varepsilon\downarrow 0}\E{\langle \frac{1}{\varepsilon}\int_{t}^{t+\varepsilon}\Et{Y_{s}}ds, \eta_{t}\rangle }.
\end{eqnarray*}
Since
\begin{eqnarray*}
\E{\left(\frac{1}{\varepsilon}\int_{t}^{t+\varepsilon}\Et{Y_{s}}ds\right)^{2}}
&\le& \E{\int_{t}^{t+\varepsilon}\frac{1}{\varepsilon^{2}}ds\int_{t}^{t+\varepsilon}\Et{Y_{s}}^{2}ds}\\
&=& \frac{1}{\varepsilon}\E{\int_{t}^{t+\varepsilon}\Et{Y_{s}}^{2}ds}\\
&\le&\frac{1}{\varepsilon}\int_{t}^{t+\varepsilon}\E{Y_{s}^{2}}ds,
\end{eqnarray*}
and $\lim_{\varepsilon\downarrow 0}\frac{1}{\varepsilon}\int_{t}^{t+\varepsilon}\E{Y_{s}^{2}}ds=\E{Y_t^2}$,
there exists a constant $\delta_t> 0$, such that
$$\E{\left(\frac{1}{\varepsilon}\int_{t}^{t+\varepsilon}\Et{Y_{s}}ds\right)^{2}}< 2\E{Y_t^2}, \quad \forall\, \varepsilon\in (0, \delta_t).$$
This implies that $\frac{1}{\varepsilon}\int_{t}^{t+\varepsilon}\Et{Y_{s}}ds$ is uniformly integrable in
$\varepsilon\in (0, \delta_t)$. Hence
$$\lim_{\varepsilon\downarrow 0}\E{\left|\frac{1}{\varepsilon}\int_{t}^{t+\varepsilon}\Et{Y_{s}}ds\right|}
=\E{\lim_{\varepsilon\downarrow 0}\left|\frac{1}{\varepsilon}\int_{t}^{t+\varepsilon}\Et{Y_{s}}ds\right|}=0.$$
Since $\eta$ is essentially bounded, so is $\eta_t$; hence there exists a constant $c>0$ such that
\begin{eqnarray*}
\left|\E{\langle \frac{1}{\varepsilon}\int_{t}^{t+\varepsilon}\Et{Y_{s}}ds, \eta_{t}\rangle }\right|
&\le& c\E{\left|\frac{1}{\varepsilon}\int_{t}^{t+\varepsilon}\Et{Y_{s}}ds\right|}\\
&\rightarrow&0,
\end{eqnarray*}
implying  $$\lim_{\varepsilon\downarrow 0}\E{\langle \frac{1}{\varepsilon}\int_{t}^{t+\varepsilon}\Et{Y_{s}}ds, \eta_{t}\rangle }=0.$$
Thus $\E{\langle Y_{t}, \eta\rangle}=0, \; a.e. t\in [0,T]$ for any $\eta \in {\mathcal D}$,  which implies
$$Y_t=0, \; a.e. t\in [0,T], \; a.s..$$
\eof

We are now in the position to present the main result of this section.

\begin{theorem}\label{maingeneral}
Given a  control $u^*\in L^2_{\cF}(0,T; \R^l)$,
let $X^{*}$ be the corresponding state process
and $(p(\cdot; t),k(\cdot;t))\in L^2_{\cF}(t,T; \R^n)\times (L^2_{\cF}(t,T; \R^n))^{d}$
be the unique solution to the BSDE (\ref{adjoint1general}).
Then $u^{*}$ is an equilibrium control if and only if
\begin{equation}\label{cond2}
\Lambda(t;t)=0, \mbox{a.s., a.e. } t\in [0, T].
\end{equation}
\end{theorem}


\pf  Recall that we have the representation $\Lambda(s; t)=\lambda_{1}(s)+\lambda_{2}(s)\xi_{t}$.
Since $\lambda_{2}$ is essentially bounded and $\xi$ is continuous, we have
\begin{eqnarray*}
\lim_{\varepsilon\downarrow 0}\Et{\frac{1}{\varepsilon}\int_{t}^{t+\varepsilon}| \lambda_{2}(s)(\xi_{s}-\xi_{t})|ds}
&\le&c \lim_{\varepsilon\downarrow 0}\frac{1}{\varepsilon}\int_{t}^{t+\varepsilon}\Et{|\xi_{s}-\xi_{t}|}ds\\
&=&0,
\end{eqnarray*}
where the last equality is because $\Et{|\xi_{s}-\xi_{t}|}$ is a continuous function of $s$ and vanishes at $s=t$.

It then follows
$$\lim_{\varepsilon\downarrow 0}\frac{1}{\varepsilon}\int_{t}^{t+\varepsilon}\Et{\Lambda(s;t)}ds=
\lim_{\varepsilon\downarrow 0}\frac{1}{\varepsilon}\int_{t}^{t+\varepsilon}\Et{\Lambda(s;s)}ds.$$

Now, if (\ref{cond2}) holds, then $$\lim_{\varepsilon\downarrow 0}\frac{1}{\varepsilon}\int_{t}^{t+\varepsilon}\Et{\Lambda(s;t)}ds
=\lim_{\varepsilon\downarrow 0}\frac{1}{\varepsilon}\int_{t}^{t+\varepsilon} \Et{\Lambda(s;s)}ds=0.$$

Conversely, if (\ref{cond1}) holds,  then  $\lim_{\varepsilon\downarrow 0}\frac{1}{\varepsilon}\int_{t}^{t+\varepsilon} \Et{\Lambda(s;s)}ds=0$, leading to
(\ref{cond2}) by virtue of
Lemma \ref{cond-lebesgue}.
\eof

\section{Uniqueness When State is One-dimensional and Coefficients Are Deterministic}\label{sec-uniq}

In our previous paper \cite{HJZ}, when the state variable is scalar-valued, i.e., $n=1$, and all
the coefficients are deterministic, an explicit equilibrium is constructed essentially based on the equivalent condition (\ref{cond2}) (although we were not yet able to prove it there). In this section, we will prove that in the same
setting the equilibrium is actually unique, thanks again to (\ref{cond2}).

Throughout this section we assume that $n=1$ and all the parameters  $A, B, b, C^j, D^j, \sigma^j, Q$ and  $R$ are deterministic function of $t$. In this case the controlled system reduces to
\begin{equation}\label{control}
dX_s=[A_sX_s+B_s'u_s+b_s]ds+ [C_sX_s+D_s u_s+\sigma_s]'dW_s;\quad X_0=x_0,
\end{equation}
where
\[ C:=(C^1,\cdots,C^d)',\; D:=((D^1)',\cdots,(D^d)')',\; \sigma:=(\sigma^1,\cdots,\sigma^d)'.\]
Accordingly, the BSDE (\ref{adjoint1general}) is simplified  to (also noting that $k(s;t)\equiv k(s)$)
\begin{equation}  \label{adjoint1}
\left\{\begin{array}{l}
dp(s;t)=-[A_sp(s;t)+C'_s k(s)+Q_sX^*_s]ds+k(s)'dW_s,\;\;s\in[t,T],\\
p(T;t)=G X^*_T- h \mathbb E_t[X^*_T]-\mu_1 X_t^*-\mu_2,
\end{array}\right.
\end{equation}
whereas the corresponding $\Lambda(s;t)$ is now in the form
$$\Lambda(s;t)= B_{s}p(s;t)+D'_{s}k(s)+R_{s} u^*_{s}.$$

In \cite{HJZ}, an equilibrium control was constructed through the solution of the following system of ODEs  (where we suppress  subscripts ${s}$ for notational simplicity):
\begin{eqnarray}
&&
\left\{\begin{array}{l}
0=\dot{M}+(2A+|C|^2)M+Q\\
\;\;-M(B'+C'D)(R+MD'D)^{-1}[(M-N-\Gamma^{(1)})B+MD'C], \;s\in[0,T],\\
M_T=G;
\end{array}\right. \label{Ric1}\\
&&\left\{\begin{array}{l}
0=\dot{N}+2AN\\
\;\;-NB'(R+MD'D)^{-1}[(M-N-\Gamma^{(1)})B+MD'C], \;s\in[0,T],\\
N_T=h;
\end{array}\right.
\label{Ric2}\\
&&\left\{\begin{array}{l}
  \dot{\Gamma}^{(1)}=-A\Gamma^{(1)},\;\;s\in[0,T],\\
  \Gamma^{(1)}_T=\mu_1 ;
  \end{array}\right.
\label{Ric3}\\
&&\left\{\begin{array}{l}
0=\dot{\Phi}+\{A-[(M-N) B'+MC'D](R+MD'D)^{-1}  B\}\Phi+(M-N)b\\
\;\; +C'M\sigma -[(M-N) B'+MC'D](R+MD'D)^{-1}MD'\sigma ,\;s\in[0,T],\\
\Phi_T=-\mu_2.
\end{array}\right.\label{Ric4}
\end{eqnarray}

If this system of equations admits a solution $(M, N, \Gamma^{(1)}, \Phi)$, then the feedback control law
\begin{equation}\label{usstar}
u^*_s=\alpha_s X^*_s+\beta_s
\end{equation}
defines an equilibrium,
where
\begin{equation}\label{ab}
\begin{array}{l}
\alpha_s\trieq  -(R_s+M_sD'_sD_s)^{-1}[(M_s-N_s-\Gamma_s^{(1)})B_s+M_sD'_sC_s],\\
\beta_s\trieq -(R_s+M_sD'_sD_s)^{-1}(\Phi_sB_s+M_sD'_s\sigma_s);
\end{array}
\end{equation}
see \cite[Theorem 4.4]{HJZ}. Moreover, the existence of solution
to (\ref{Ric1})--(\ref{Ric4}) is studied in \cite{HJZ}.

The next theorem provides  that the control constructed above is the {\it only} equilibrium.
\begin{theorem}
If (\ref{Ric1})--(\ref{Ric4}) admits a solution $(M, N, \Gamma^{(1)}, \Phi)$, then
there is a unique equilibrium control.
\end{theorem}
\pf  Suppose there is another equilibrium state--control pair $(X, u)$. Then, with a slight abuse of notation,  equation (\ref{adjoint1general}),  with $X^{*}$ replaced by $X$,
 admits a unique solution $(p(\cdot;t), k(\cdot))$ satisfying $\Lambda(s;s)\equiv B_{s}p(s;s)+D_{s}'k(s)+R_{s}u_{s}=0$ for a.e. $s\in [0, T]$.

Define
$$\begin{array}{l}
\bar p(s;t):=p(s;t)-(M_{s}X_{s}-N_{s}\Et{X_{s}}-\Gamma^{(1)}_{s}X_{t}+\Phi_{s}), \\
 \bar k(s):=k(s)-M_{s}(C_{s}X_{s}+D_{s}u_{s}+\sigma_{s}).
 \end{array}
 $$

The equilibrium condition for $(X,u)$ yields
$$B_{s}\left[\bar p(s;s)+(M_{s}-N_{s}-\Gamma^{(1)}_{s})X_{s}+\Phi_{s}\right]
+D_{s}'\left[\bar k(s) +M_{s}(C_{s}X_{s}+D_{s}u_{s}+\sigma_{s})\right]+R_{s}u_{s}=0.
$$
Since $R_{s}+D'_{s}M_{s}D_{s}$ is invertible, we solve for $u_s$ in the above equation to obtain the following expression
\begin{equation}\label{us}
\begin{array}{rl}
u_{s}=&-(R_{s}+D'_{s}M_{s}D_{s})^{-1}\left[B_{s}\bar p(s;s)+D_{s}'\bar k(s)\right.\\
&\;\;\left.+(B_{s}(M_{s}-N_{s}-\Gamma^{(1)}_{s})+D_{s}'M_{s}C_{s})X_{s}
+B_{s}\Phi_{s}+D_{s}'M_{s}\sigma_{s}\right].
\end{array}
\end{equation}
On the other hand, we can show that $(\bar p(\cdot;t), \bar k(\cdot))$ satisfies the following BSDE (details are relegated to Appendix A):
\begin{equation}\label{dpbar}
\left\{\begin{array}{rl}
d\bar p(s;t)
=&-\left(A\bar p(s;t)+C'\bar k(s) -[C'MD+MB'][R+D'MD]^{-1}[B\bar p(s;s)+D'\bar k(s) ]\right.\\
  &\;\;\left. +NB' [R+D'MD]^{-1}\Et{B\bar p(s;s)+D'\bar k(s) }\right)ds+\bar k(s)' dW_{s},\;\;s\in [t,T],\\
  \bar p(T;t)=0,
  \end{array}\right.
\end{equation}
where we suppress the subscript ${s}$ for $A,B,C,D,M,N,R$, and  we have used the equations for $M, N, \Gamma^{(1)}, \Phi$. Moreover, it is easy to prove that
$\E{\int_{0}^{T}|\bar k(s)|^{2}ds}<+\infty$ and  $\sup_{t\in [0, T]}\E{\sup_{s\ge t}|\bar p(s;t)|^{2}}<+\infty.$

We will prove in the next theorem that equation (\ref{dpbar})  admits at most one solution in the space ${\mathcal L}_{1}\times {\mathcal L}_{2}$, where

$${\mathcal L}_{1}:=\left\{X(\cdot;\cdot):  X(\cdot;t)\in L^{2}_{\cF}(t, T; \R), \sup_{t\in [0, T]}\E{\sup_{s\ge t}|X(s;t)|^{2}}<+\infty\right\},$$
and
$${\mathcal L}_{2}:= \left\{Y(\cdot;\cdot): Y(\cdot;t)\in L^{2}_{\cF}(t,T; \R^{d}), \sup_{t\in [0, T]}\E{\int_{t}^{T}|Y(s;t)|^{2}ds}<+\infty\right\}.$$
Hence
$\bar p(s;t)\equiv 0$ and $\bar k(s)\equiv 0$.

Finally, plugging $\bar p\equiv \bar k\equiv 0$ into (\ref{us}), we find that
$u$ has exactly the same form of
feedback control as that of $u^*$; see (\ref{usstar}). This proves that $u$ and $u^*$ lead to an identical control.
\eof

It remains to prove the uniqueness of solution for (\ref{dpbar}). Indeed we will do it for a
 more  general equation
\begin{equation}\label{subtriangle}
\left\{\begin{array}{rl}
d\bar p(s;t)=&-f\left(s, \bar p(s;t),  \bar p(s;s), \Et{ l_{1}(s) \bar p(s;s)},\bar k(s;t) ,\Et{l_{2}(s)\bar k(s;t) }\right)ds\\
&\;\;\;\;+\bar k(s;t)' dW_{s},\;\;s\in[t,T],\\
\bar p(T;t)=&0,
\end{array}\right.
\end{equation}
 where $l_{1}$ and $l_2$ are two essentially bounded, adapted vector processes with suitable dimensions, 
 and $f(s, \cdots\cdot\cdot)$ is a deterministic function satisfying uniform Lipschitz condition in all variables except $s$.

 \begin{theorem}\label{uni-subtriangle}
 Equation (\ref{subtriangle}) admits at most one solution $(\bar p, \bar k)$ in the space ${\mathcal L}_{1}\times {\mathcal L}_{2}$.
 \end{theorem}
\pf
Suppose there are two solutions $(\bar p^{(1)}, \bar k^{(1)})$ and $(\bar p^{(2)}, \bar k^{(2)})$ in the space ${\mathcal L}_{1}\times {\mathcal L}_{2}$. Define
$\bar p(s;t)\trieq \bar p^{(1)}(s;t)-\bar p^{(2)}(s;t), \bar k(s;t)\trieq \bar k^{(1)}(s;t)-\bar k^{(2)}(s;t)$ and
\begin{eqnarray*}
\Delta f(s;t)&\trieq& f(s, \bar p^{(1)}(s;t),  \bar p^{(1)}(s;s), \Et{ l_{1}(s) \bar p^{(1)}(s;s)}, \bar k^{(1)}(s;t) ,\Et{l_{2}(s)\bar k^{(1)}(s;t) })\\
&&-f(s, \bar p^{(2)}(s;t),  \bar p^{(2)}(s;s), \Et{ l_{1}(s) \bar p^{(2)}(s;s)}, \bar k^{(2)}(s;t) ,\Et{l_{2}(s)\bar k^{(2)}(s;t) }).
\end{eqnarray*}
Then $|\Delta f(s;t)|\le c_{1} \left( |\bar p(s;t)|+|\bar k(s;t)|+|\bar p(s;s)|+\Et{|\bar p(s;s)|}+\Et{ |\bar k(s;t)|}\right)$ for some constant $c_{1}$, and
$$d\bar p(s;t)=-\Delta f(s;t)dt+\bar k(s;t)'dW_{s}, \quad \bar p(T;t)=0.$$

 For any $t\in [0,T]$, $s\in [t, T]$, by It\^o's formula, we have
$$|\bar p(s;t)|^{2}+\int_{s}^{T}|\bar k(u;t)|^{2}du
=2\int_{s}^{T}\bar p(u;t) \Delta f(u;t)du-2\int_{s}^{T}\bar p(u;t)\bar k(u;t)'dW_{u}.
$$
Thus  
\begin{eqnarray*}
&&\E{|\bar p(s;t)|^{2}}+\E{\int_{s}^{T}|\bar k(u;t)|^{2}du} \\
&\le& c_{1}\E{\int_{s}^{T} |\bar p(u;t)| \left( |\bar p(u;t)|+|\bar k(u;t)|+|\bar p(u;u)|+\Et{|\bar p(u;u)|}+\Et{ |\bar k(u;t)|}\right)du} \\
&\le& c_{2}\E{\int_{s}^{T} \left( |\bar p(u;t)|^{2}+|\bar p(u;u)|^{2}\right)du}  +\frac{1}{2} \E{\int_{s}^{T}|\bar k(u;t)|^{2}du},
\end{eqnarray*}
where we have used the inequality $cxy\le c^{2}x^{2}+\frac{1}{4}y^{2}$ for any nonnegative $c,x,y$.
Consequently,  there exists $c_{3}>0$ such that
\begin{equation}\label{foralline}
\E{|\bar p(s;t)|^{2}}+\E{\int_{s}^{T}|\bar k(u;t)|^{2}du}\le c_{3}\E{\int_{s}^{T} \left( |\bar p(u;t)|^{2}+|\bar p(u;u)|^{2}\right)du}  .
\end{equation}
Furthermore, for any $s\in [t, T]$, we have
\begin{eqnarray*}
\E{|\bar p(s;t)|^{2}+\int_{s}^{T}|\bar k(u;t)|^{2}du}
&\le& c_{3}(T-t)\left[ \sup_{u\in [t, T]}\E{ |\bar p(u;t)|^{2}} +  \sup_{u\in [t, T]}\E{|\bar p(u;u)|^{2}}\right ] \\
&\le& 2c_{3}(T-t)\sup_{t\le u\le s\le T} \E{|\bar p(s;u)|^{2}}.
\end{eqnarray*}
Hence
\begin{equation}\label{fortriangle}
\sup_{t\le u\le s\le T} \E{|\bar p(s;u)|^{2}}\le  2c_{3}(T-t)\sup_{t\le u\le s\le T} \E{|\bar p(s;u)|^{2}}.
\end{equation}

Now take  $\delta\in (0, 1/(4c_3))$. Then
for any $t\in [T-\delta, T]$, we have
\begin{eqnarray*}
\sup_{t\le u\le s\le T} \E{|\bar p(s;u)|^{2}}&\le& \frac{1}{2}\sup_{t\le u\le s\le T} \E{|\bar p(s;u)|^{2}},
\end{eqnarray*}
which implies $\sup_{t\le u\le s\le T} \E{|\bar p(s;u)|^{2}}=0$. It follows that $\bar p(s;u)=0, a.s.$ almost everywhere
in $\{(s,u): t\le u\le s\le T\}$.

For $t\in [T-2\delta, T-\delta]$ and $s\in [T-\delta, T]$, since $\bar p(u,u)=0$ for any $u\in [s, T]$,
we have by (\ref{foralline}) that
\begin{equation}
\E{|\bar p(s;t)|^{2}}+\E{\int_{s}^{T}|\bar k(u;t)|^{2}du}\le c_{3}\E{\int_{s}^{T} |\bar p(u;t)|^{2}du}  .
\end{equation}
Grownwall's inequality then leads to $\bar p(s;t)=\bar k(s;t)=0$.

For $t\in [T-2\delta, T-\delta]$ and $s\in [t, T-\delta]$, noting $\bar p(T-\delta;t)=0$, we can apply the previous argument for the region $t\in [T-\delta, T]$ and $s\in [t, T]$ to deduce that $\bar p(s;t)=\bar k(s;t)= 0$.

We can then repeat the same analysis in  a backward manner to $t\in [T-3\delta, T-2\delta]$ and so on until we reach time $t=0$.
\eof

\section{Uniqueness of Mean-Variance Equilibrium Strategies in A Complete Market with Random Parameters}\label{mvcase}

Following \cite{HJZ}, as an application of the time-inconsistent LQ theory, we study the continuous-time Markowitz mean--variance portfolio selection
model in a complete market with random model coefficients. We aim to establish the uniqueness of the equilibrium strategy.
The model is mathematically a special case of the general LQ problem formulated
earlier in this paper, with $n = 1$ naturally. However, since some coefficients are allowed to
be random, the uniqueness result of the previous section is not applicable here.

We use the same setup of \cite{HJZ}. The wealth equation is governed by the SDE
\begin{equation}\label{wealth2}
 \left\{\begin{array}{l}
         dX_s=r_s X_sds+\theta_s'u_sds+u_s'dW_s,\qquad s\in [t,T],\\
     X_t=x_t,
        \end{array}
 \right.
\end{equation}
where $r$ is the (bounded) deterministic interest rate function, and $\theta$ is the essentially bounded stochastic
 risk premium process.

The objective at time $t$ with state $X_{t}=x_{t}$ is to minimize
\begin{eqnarray}\label{mv-obj}
J(t, x_t; u)&\trieq &\frac{1}{2}{\rm Var}_t(X_T)-(\mu_1 x_t+\mu_2) \mathbb E_t[X_T]\\
&=&\frac{1}{2}\left(\mathbb E_t[X_T^2]-(\mathbb E_t[X_T])^2\right)-(\mu_1 x_t+\mu_2) \mathbb E_t[X_T]\nonumber
\end{eqnarray}
with $\mu_1\ge 0$. As noted in \cite{HJZ}, there are two sources of time-inconsistency in this model, one from the variance term and the other from the state-dependent tradeoff between
the mean and the variance.

In \cite[Section 5]{HJZ}, we constructed an equilibrium through the solutions
$(M,U)$,  $(\Gamma^{(1)}, \gamma^{(1)})$,  $(\Gamma^{(2)}, \gamma^{(2)})$, and $(\Gamma^{(3)}, \gamma^{(3)})$
to BSDEs:
\begin{equation}\label{5bsde}
\left\{
\begin{array}{rcl}
dM_s&=&-[2r_sM_s+(\theta_s M_s+U_s)'\alpha_s] ds+U_s 'dW_s,\quad M_T=1,\\
d\Gamma^{(1)}_s&=&-r_s\Gamma^{(1)}_sds+(\gamma^{(1)}_s)' dW_s,\quad \Gamma^{(1)}_T=\mu_1,\\
d\Gamma^{(2)}_s&=&-[r_s\Gamma^{(2)}_s+(\theta_s M_s+U_s)'\beta_s] ds+(\gamma^{(2)}_s)'dW_s,\quad \Gamma^{(2)}_T=-\mu_2,\\
d\Gamma^{(3)}_s&=&-[r_s\Gamma^{(3)}_s + (\theta_s M_s+U_s)'\beta_s]ds+(\gamma^{(3)}_s)'dW_s,\quad \Gamma^{(3)}_T=0,
\end{array}\right.
\end{equation}
where
\begin{equation}\label{abmv}
\begin{array}{l}
\alpha_s\trieq -M_s^{-1}\left(-\theta_s \Gamma^{(1)}_s+U_s-\gamma^{(1)}_s\right),\\
 \beta_s\trieq -M_s^{-1}\left[\theta_s (\Gamma^{(2)}_s-\Gamma^{(3)}_s)+\gamma^{(2)}_s\right].
\end{array}
\end{equation}

 In this case, the BSDE (\ref{adjoint1general}) for $p(\cdot;t)$ corresponding to a given strategy (control) $u^*$ with the wealth (state) process $X^*$ specializes to
 \begin{equation}\label{adjointmv}
 \left\{\begin{array}{l}
 dp(s;t)=-r_s p(s;t)ds+k(s)'dW_s,\\
p(T;t)=X^*_T-\mathbb E_t[X^*_T]-\mu_1X^*_t-\mu_2,
\end{array}\right.
\end{equation}
and the corresponding $\Lambda(s;t)$ is
$$\Lambda(s;t)=p(s;t)\theta_s+k(s).$$

It is proved in \cite[Proposition 5.1]{HJZ} that the system of BSDEs (\ref{5bsde}) admits a unique solution with  both $M$ and  $M^{-1}$ being  {\it bounded}, and $U\cdot W$  a BMO martingale. Furthermore, the
feedback strategy
\begin{equation}\label{mvequal}
u^*_s=\alpha_sX_s^*+\beta_s
\end{equation}
defines a  control in the space $L^{2}_{\cF}(0,T; \R^{d})$, which is an equilibrium strategy for the mean--variance investment problem.

We now claim that the equilibrium above is unique.

For any $q>1$, define
$${\mathcal L}_{3}(q):=\left\{X(\cdot;\cdot): X(\cdot;t)\in L^{q}_{\cF}(\Omega; C(t, T; \R))\;\; \forall\, t\in [0, T]\right\},$$
and
$${\mathcal L}_{4}(q):= \left\{Y(\cdot): Y \mbox{ is adapted and }   \E{\left(\int_{t}^{T}|Y(s)|^{2}ds\right)^{q/2}}<+\infty\right\}.$$

\begin{theorem}
There is a unique equilibrium strategy for the mean--variance problem (\ref{wealth2})--(\ref{mv-obj}), which is identical to the one generated from the
feedback law  (\ref{mvequal}).
\end{theorem}
\pf  Suppose there is another equilibrium wealth--strategy pair $(X, u)$. Then  equation (\ref{adjointmv}),  with $X^{*}$ replaced by $X$,
 admits a unique solution $(p(\cdot;t), k(\cdot))$ satisfying $\Lambda(s;s)\equiv p(s;s)\theta_{s}+k(s)=0$ for a.e. $s\in [0, T]$.


It is proved in  \cite{HJZ} that $M, M^{-1}$, $\Gamma^{(1)}$, $\Gamma^{(2)}$ and $\Gamma^{(3)}$ are all bounded,  and $ \gamma^{(2)}\cdot W$ and  $U\cdot W$ are both BMO martingales.
In particular, since  $U\cdot W$ is a BMO martingale, it follows from the John--Nirenberg inequality (see Kazamaki \cite[Theorem 2.2, p.29]{kazamaki}) that there exists $\varepsilon>0$ such that $\E{e^{\varepsilon \int_{0}^{T}|U_{s}|^{2}ds}}<+\infty$. Thus $\E{\left(\int_{0}^{T}|U_{s}|^{2}ds\right)^{q}}<+\infty$ for any $q>0$.

Define
\[\begin{array}{l}
\bar p(s;t):=p(s;t)-\left[M_{s}X_{s}+\Gamma^{(2)}_{s}-\mathbb{E} \left(M_{s}X_{s}+\Gamma^{(3)}_{s}\right)
-\Gamma^{(1)}_{s}X_{t}\right],\\
\bar k(s)=k(s)-\left(M_{s}u_{s}+U_{s}X_{s}+\gamma^{(2)}_{s}\right).
\end{array}
\]
It is easy to check that $\bar p\in {\mathcal L}_{3}(2)$. On the other hand,
 $k\in L^{2}_{\cF}(0,T;\R^{d}), Mu+\gamma^{(2)}\in L^{2}_{\cF}(0,T;\R^{d})$,
and for  any $ q\in(1,2)$,
\begin{eqnarray*}
\E{\left(\int_{0}^{T}|U_{s}X_{s}|^{2}ds\right)^{q/2}}
&\le &\E{\sup_{s\in [0, T]}|X_{s}|^{q}\left(\int_{0}^{T}|U_{s}|^{2} ds\right)^{q/2}}\\
&\le& \left(\E{\sup_{s\in [0, T]}|X_{s}|^{2}}\right)^{q/2}\left(\E{\left(\int_{0}^{T}|U_{s}|^{2}ds\right)^{q/(2-q)}}\right)^{1-q/2}\\
&<&+\infty.
\end{eqnarray*}
These, together with the fact that $L^{2}_{\cF}(0,T; \R^{d})\subset {\mathcal L}_{4}(q)$ $\forall q\in (1, 2)$, imply
  $\bar k\in {\mathcal L}_{4}(q)$ for $q\in (1, 2)$.

Furthermore, the equivalent condition gives
\[
\bar p(s;s)\theta_s +\bar k(s) +\theta_s[\Gamma^{(2)}_{s}-\Gamma^{(3)}_{s}-\Gamma^{(1)}_{s}X_{s}]
+[M_{s}u_{s}+U_{s}X_{s}+\gamma^{(2)}_{s}]=0.
\]
Solving for $u_s$ we obtain
\begin{equation}\label{mvus}
\begin{array}{rl}
u_{s}=&-M_{s}^{-1}\left[(U_{s}-\theta_s\Gamma^{(1)}_{s})X_{s}+\theta_s\bar p(s;s)+\bar k(s)+\theta_s(\Gamma^{(2)}_{s}-\Gamma^{(3)}_{s})+\gamma^{(2)}_{s}\right]\\
=&\alpha_{s}X_{s}+\beta_{s}-M_{s}^{-1}[\theta_s\bar p(s;s)+\bar k(s)].
\end{array}
\end{equation}

Next, we can derive the following BSDE that $(\bar p(\cdot;t),\bar k(\cdot))$ satisfies (details are placed in 
Appendix B)
\begin{equation}\label{barp0}
\left\{\begin{array}{ll}
d\bar p(s;t)=&-\left\{ r_{s}\bar p(s;t)-(\theta_{s}+U_{s}M_{s}^{-1})'[\theta_{s}\bar p(s;s)+\bar k(s)]\right.\\
&\left.+\Et{(\theta_{s}+U_{s}M_{s}^{-1})'[\theta_{s}\bar p(s;s)+\bar k(s)]}\right\}ds
+\bar k(s)'dW_{s},\;s\in [t,T],\\
\bar p(T;t)=&0.
\end{array}\right.
\end{equation}

We will prove in the next theorem that this equation admits at most one solution $(\bar p, \bar k)$ in the space
${\mathcal L}_{3}(q) \times {\mathcal L}_{4}(q)$ for some $q\in (1,2)$,
leading to $\bar p\equiv 0$ and $\bar k\equiv 0$. Consequently,  
we have $u_{s}=\alpha_{s}X_{s}+\beta_{s}$. In other words, $u_s$ has exactly the same feedback form as $u^*_s$. This establishes the uniqueness. \eof

\begin{theorem}
 For any $q\in (1,2)$, equation (\ref{barp0}) admits at most one solution $(\bar p, \bar k)\in {\mathcal L}_{3}(q)\times {\mathcal L}_{4}(q)$.
\end{theorem}
\pf
Fix $t$. Taking  $\Et{\cdot}$ on both sides of  the integral form of
(\ref{barp0}) and noticing that $\int_t^s\bar k\cdot W$ is a martingale, we get
$$\Et{\bar p(s;t)}=\int_{s}^{T} r_{\nu}\Et{\bar p(\nu;t)}d\nu, $$
which implies $\Et{\bar p(s;t)}=0$ for any $s\ge t$. In particular, taking $s=t$, we have $\bar p(t;t)=0$. Hence equation (\ref{barp0}) reduces to
 \begin{equation}\label{barp1}
\left\{\begin{array}{l}
d\bar p(s;t)=-\left\{ r_{s}\bar p(s;t)-(\theta_{s}+U_{s}M_{s}^{-1})'\bar k(s)+\Et{(\theta_{s}+U_{s}M_{s}^{-1})'\bar k(s)}\right\}ds+\bar k(s)'dW_{s},\\
\bar p(T;t)=0.
\end{array}\right.
\end{equation}

As $r$ is deterministic and bounded, we can discount $\bar p(s;t)$ by $e^{-\int_{s}^{T}r_{v}dv}$ to remove the term $-r_{s}\bar p(s;t)$ on the right hand side of the above equation; thus henceforth we  assume $r\equiv 0$ without loss of generality.  Define
$\tilde p(s;t):=\bar p(s;t)-\int_{s}^{T}\Et{(\theta_{v}+U_{v}M_{v}^{-1})'\bar k(v)}dv$. Then  $\tilde p(T;t)=0$ and
$$d\tilde p(s;t) = (\theta_{s}+U_{s}M_{s}^{-1})'\bar k(s)ds+\bar k(s)'dW_{s}.$$

For any $\bar q\in (1, q)$, denote $\hat q=q/\bar q$, and $1/\hat p+1/\hat q=1$. Then
\begin{eqnarray*}
&&\E{\sup_{s\in [t,T]}\left| \int_{s}^{T}\Et{\left( \theta_{\nu}+U_{\nu}M_{\nu}^{-1}\right)  '\bar k\left( \nu\right)   }d\nu\right|^{\bar q}}\\
&\le&\E{\left(\int_{t}^{T}\left|\left( \theta_{\nu}+U_{\nu}M_{\nu}^{-1}\right)  '\bar k(\nu)\right| d\nu\right)^{\bar q}}\\
&\le&c_{0}\E{\left( \int_{t}^{T}|\theta_{\nu}'\bar k\left( \nu\right)  | d\nu\right)  ^{\bar q}}+c_{0}\E{\left( \int_{t}^{T} M_{\nu}^{-1}|U_{\nu}'\bar k\left( \nu\right)  |d\nu \right)  ^{\bar q} }\\
&\le&c_{1}\E{\left( \int_{t}^{T}|\bar k\left( \nu\right)  |^{2}d\nu \right)  ^{\bar q/2}} +c_{2}\E{\left( \int_{t}^{T}|U_{\nu}|^{2}d\nu\right)^{\bar q/2}\left(\int_{t}^{T}|\bar k(\nu)|^{2}d \nu\right)^{\bar q/2}}\\
&\le &c_{3} +c_{2}\left(\E{\left( \int_{t}^{T}|U_{\nu}|^{2}d\nu\right)^{\bar q\hat p /2}}\right)^{1/\hat p}
\left(\E{\left(\int_{t}^{T} |\bar k(\nu)|^{2}d \nu\right)^{q/2} }\right)^{1/\hat q}\\
&<&+\infty,
\end{eqnarray*}
where $c_{0}, c_{1}, c_{2}$ and $ c_{3}$ are proper constants.
On the other hand, it is assumed that $\bar p\in {\mathcal L}_{3}(q)$. So it follows that
$\E{\sup_{s\in [t, T]}|\tilde p(s;t)|^{\bar q}}<+\infty$.

Define $\xi={\mathcal E}(-(\theta_{s}+U_{s}M_{s}^{-1}) \cdot  W)_{T}\equiv e^{-\frac{1}{2}\int_{0}^{T}|\theta_{s}+U_{s}M_{s}^{-1}|^{2}ds-\int_{0}^{T}(\theta_{s}+U_{s}M_{s}^{-1})'dW_{s}}$. Since $UM^{-1}\cdot W$ is a BMO martingale, $\E{\xi}=1$; so it can be used to define  a new probability measure $\Q$ by
$\frac{d\Q}{d{\mathbb P}}=\xi$, under which $\hat W_{s}=W_{s}+\int_{0}^{s}(\theta_{v}+U_{v}M_{v}^{-1})dv$ is a standard Brownian motion. Furthermore,
$$d\tilde p(s;t)=\bar k(s)'d\hat W_{s},\;\; \tilde p(T;t)=0.$$

Applying It\^o's formula, we obtain
\begin{eqnarray*}
dM^{-1}_s&=& -M^{-2}_{s}dM_{s}+M^{-3}_{s}U_{s}^{2}ds\\
&=&M^{-1}_{s}\left\{ \left[\theta(\frac{\Gamma^{(1)}_{s}}{M_{s}}-1) \frac{U_{s}}{M_{s}}+\frac{\Gamma^{(1)}_{s}|\theta_{s}|^{2}}{M}\right]ds -\frac{U_{s}'}{M_{s}}dW_{s}\right\}.
\end{eqnarray*}
Hence
$$M^{-1}_{T}=M^{-1}_{0}\exp\left(-\int_{0}^{T}  \left[\frac{U_s'\theta_s}{M_s}-\Gamma^{(1)}_s
\frac{|\theta_s|^2}{M_s}+\frac{1}{2}\frac{|U_s|^2}{M_s^2}-\Gamma^{(1)}_s\frac{U_s'\theta_s}{M_s^2}\right]ds
-\int_{0}^{T}\frac{U'_{s}}{M_{s}}dW_{s}\right).
$$
Comparing $\xi$ and $M^{-1}_{T}$, we deduce
$$\xi M_{T}=M_{0}\exp\left(-\int_{0}^{T}\Gamma^{(1)}_{s}|\theta_{s}|^{2}\frac{1}{M_{s}}ds\right)
\exp\left(-\int_{0}^{T}\Gamma^{(1)}_{s}\frac{\theta_{s}'}{M_{s}}\frac{U_{s}}{M_{s}}ds\right) \exp\left(-\frac{1}{2}\int_{0}^{T} |\theta_{s}|^{2}ds-\int_{0}^{T}\theta_{s}dW_{s}\right).$$
It is clear that $M_{0}e^{-\int_{0}^{T}\Gamma^{(1)}_{s}|\theta_{s}|^{2}\frac{1}{M_{s}}ds}$ is bounded, and
$ e^{-\frac{1}{2}\int_{0}^{T} |\theta_{s}|^{2}ds-\int_{0}^{T}\theta_{s}'dW_{s}}\in L^{\bar q}$ for any $\bar q>1$. Moreover,
for any $\bar q>1$ and any $\varepsilon>0$, there exists a constant $C>0$ such that
$$\E{\left(e^{-\int_{0}^{T}\Gamma^{(1)}_{s}\frac{\theta_{s}}{M_{s}}\frac{U_{s}}{M_{s}}ds} \right)^{\bar q} }
\le C\E{e^{\varepsilon \int_{0}^{T}|U_{s}|^{2}ds}}.$$
We have shown previously that  there exists $\varepsilon>0$ such that $\E{e^{\varepsilon \int_{0}^{T}|U_{s}|^{2}ds}}<+\infty$. Therefore
$e^{-\int_{0}^{T}\Gamma^{(1)}_{s}\frac{\theta_{s}}{M_{s}}\frac{U_{s}}{M_{s}}ds} \in L^{\bar q}$.
This in turn proves $\xi M_{T}\in L^{\bar q}$. However, $M^{-1}$ is  bounded, so
 $\xi\in L^{\bar q}$ for any $\bar q>1$.

Now for any  $\bar q\in (1,q)$ and $\hat q\in (1, \bar q)$,   we have
\begin{eqnarray*}
{\mathbb E}^{\Q}[\sup_{s\in [t, T]}|\tilde p(s;t)|^{\hat q}]&=&\E{\sup_{s\in [t, T]}|\tilde p(s;t)|^{\hat q}\xi}\\
&\le& \left(\E{\sup_{s\in [t, T]}|\tilde p(s;t)|^{\bar q}}\right)^{\hat q/\bar q}\left(\E{\xi^{\bar q/(\bar q-\hat q)}}\right)^{(\bar q-\hat q)/\bar q}\\
&<&+\infty,
\end{eqnarray*}
 which implies that  $\tilde p(\cdot;t)$ is a $\Q$-martingale, and hence $\tilde p\equiv 0$ and $\bar k\equiv 0$.
 Since $\bar p(s;t)=\tilde p(s;t)+\int_{s}^{T}\Et{(\theta_{v}+U_{v}M_{v}^{-1})'\bar k(v)}dv$, we conclude
 $\bar p\equiv 0$.

\eof

\section{Concluding Remarks}\label{concl}

Equilibrium control is an alternative and weak notion of solution to a dynamic control problem when
the traditional time-consistency is absent. The uniqueness results we establish in this paper (if only for some special cases) justify, from an important aspect, not only the game formulation for the time-inconsistent dynamic decision making, but also our definition of
equilibria over the set of open-loop (instead of feedback) controls. They also shed a light on the search of conditions for uniqueness of more general problems.

Since equilibria are defined via local perturbation for the game formulation, unlike the optimal solution for a time-consistent problem, they do not inherently lead to the same value process.
The uniqueness of the solution does indeed imply the uniqueness of the value process, which
in turn addresses concerns such as ``why an equilibrium is defined  this way'', or ``which one to choose if there are multiple solutions''.

We realize that in this paper the uniqueness has been established only for some special classes of
the LQ control problem. For general time-inconsistent LQ or even non-LQ problems, existence and uniqueness of equilibria remain outstanding research problems.

 \newpage
 \appendix
 \section{Derivation of (\ref{dpbar})}

 We write (\ref{us}) as $u_{s}=\alpha_{s}X_{s}+\beta_{s}+V_{s}$ where
$V_{s}:= -(R_{s}+M_{s}D_{s}'D_{s})^{-1}[B_{s}\bar p(s;s)+D_{s}'\bar k(s)]$
and $\alpha_s$ and $\beta_s$ are given by (\ref{ab}).
The equations for $M, N, \Gamma^{(1)},\Phi$ can be rewritten as
\begin{eqnarray}
&&
0=\dot{M}+(2A+|C|^2)M+Q+M(B'+C'D)\alpha, \;s\in[0,T],\qquad
M_T=G;  \label{2Ric1}\\
&&0=\dot{N}+2AN+NB'\alpha, \;s\in[0,T],\qquad N_T=h;  \label{2Ric2}\\
&&  \dot{\Gamma}^{(1)}=-A\Gamma^{(1)},\;\;s\in[0,T],\qquad   \Gamma^{(1)}_T=\mu_1; \label{2Ric3}\\
&&\left\{\begin{array}{l}
0=\dot{\Phi}+A\Phi+[(M-N) B'+MC'D]\beta+(M-N)b+C'M\sigma  ,\;s\in[0,T],\\
\Phi_T=-\mu_2.
\end{array}\right.\label{2Ric4}
\end{eqnarray}
Hence (the subscript $s$ is suppressed)
\begin{eqnarray*}
d(MX)&=&[M(AX+B'u+b)-XQ-XM(2A+C^{2}+(B'+C'D)\alpha)]ds\\
&&+M(CX+Du+\sigma)'dW_{s}\\
&=&\left[M(B'\beta+B'V+b)-XQ-XM(A-B'\alpha+C^{2}+(B'+C'D)\alpha)\right]ds\\
&&+M(CX+Du+\sigma)'dW_{s}\\
&=&\left[M(B'\beta+B'V+b)-XQ-XM(A+C^{2}+C'D\alpha)\right]ds\\
&&+M(CX+Du+\sigma)'dW_{s}.
\end{eqnarray*}
Similarly,
\begin{eqnarray*}
d(N\Et{X_{s}})&=&[N\Et{AX+B'u+b}-N(2A+B'\alpha)\Et{X_{s}}]ds\\
&=&[N(B'\beta+B'\Et{V_{s}}+b)-N(A-B'\alpha+B'\alpha)\Et{X_{s}}]ds\\
&=&[N(B'\beta+B'\Et{V}+b)-NA\Et{X_{s}}]ds;\\
d(\Gamma^{(1)}_{s}X_{t})&=&-A\Gamma^{(1)}_{s}X_{t}ds.
\end{eqnarray*}
So
$$d(MX-N\Et{X_{s}}-\Gamma^{(1)}X_{t}+\Phi)=\zeta^{(1)}ds+(\zeta^{(2)})'dW_{s}$$
where
$\zeta^{(2)}=M(CX+Du+\sigma)$ and
\begin{eqnarray*}
\zeta^{(1)}&=&M(B'\beta+B'V+b)-XQ-XM(A+C^{2}+C'D\alpha)\\
&&-N(B'\beta+B'\Et{V}+b)+NA\Et{X_{s}}\\
&&+A\Gamma^{(1)}_{s}X_{t}\\
&&-A\Phi-[(M-N) B'+MC'D]\beta-(M-N)b-C'M\sigma\\
&=&[-Q-M(A+C^{2}+C'D\alpha)]X\;\;+NA\Et{X_{s}}\;\; +A\Gamma^{(1)}_{s}X_{t}\\
&&+(MB'V-NB'\Et{V_{s}})-A\Phi-MC'(D\beta+\sigma).
\end{eqnarray*}
However,  $\bar p(s;t)=p(s;t)-[M_{s}X_{s}-N_{s}\Et{X_{s}}-\Gamma^{(1)}_{s}X_{t}+\Phi_{s}] $, we deduce
\begin{eqnarray*}
d\bar p&=&dp-\zeta^{(1)}ds-(\zeta^{(2)})'dW_{s}\\
&=&-[A_{s}p(s;t)+C_{s}'k_{s}+Q_{s}X_{s}+\zeta^{(1)}_{s}]ds +[k_{s}-\zeta^{(2)}]'dW_{s}\\
&=&\zeta^{(3)}_{s}ds+\bar k_{s}'dW_{s},
\end{eqnarray*}
where
\begin{eqnarray*}
\zeta^{(3)}&=&-A_{s}[p(s;t)-MX+N\Et{X_{s}}+\Gamma^{(1)}_{s}X_{t}-\Phi] \\
&&-C_{s}'(k_{s}-CMX-MD\alpha X-MD\beta-M\sigma)\\
&&        -(MB'V-NB'\Et{V_{s}}) \\
&=&-A_{s}\bar p(s;t)-C'\bar k_{s}  -C'MDV-(MB'V-NB'\Et{V_{s}})\\
&=&-A_{s}\bar p(s;t)-C'\bar k_{s}  -(C'MD+MB')V+NB'\Et{V_{s}}).
\end{eqnarray*}
This proves (\ref{dpbar}).

\section{Derivation of (\ref{barp0})}
We write (\ref{mvus}) as
$u_{s}=\alpha_{s}X_{s}+\beta_{s}+V_{s},$
where $V_{s}:=-M_{s}^{-1}[\theta_{s}\bar p(s;s)+\bar k(s)]$ and $\alpha_s$ and $\beta_s$ are given by (\ref{abmv}).

Making use of (\ref{5bsde}), we can compute
\begin{eqnarray*}
d[MX]&=&[M(rX+\theta'u)-X(2rM+(M\theta+U)'\alpha)+u'U]ds+[Mu+XU]'dW_{s}\\
&=&[-rMX+(\theta M+U)'(\beta+V)]ds+[Mu+XU]'dW_{s};\\
d\Et{MX}&=&\Et{-rMX+(\theta M+U)'(\beta+V)}ds;\\
d\Gamma^{(1)}_{s}X_{t}&=&-r\Gamma^{(1)}_{s}X_{t}ds,
\end{eqnarray*}
where we have used the fact that  $\gamma^{(1)}\equiv 0$, which can be seen from the BSDE for $\Gamma^{(1)}$.

Since
$$\bar p(s;t)=p(s;t)-M_{s}X_{s}-\Gamma^{(2)}_{s}+\Et{M_{s}X_{s}+\Gamma^{(3)}_{s}}+\Gamma^{(1)}_{s}X_{t}, \;
\bar k(s)=k(s)-M_{s}u_{s}-X_{s}U_{s}-\gamma^{(2)}_{s},$$
we derive $d\bar p(s;t)=\zeta^{(4)}_{s}ds+(\zeta^{(5)})_{s}'dW_{s}$, where
$\zeta^{(5)}=k(s)-[M_{s}u_{s}+X_{s}U_{s}+\gamma^{(2)}]$, and
\begin{eqnarray*}
\zeta^{(4)}&=&-rp(s;t)\;\;\; -[-rMX+(\theta M+U)'(\beta+V)]\;\; \;+[r\Gamma^{(2)}+(\theta M+U)'\beta] \\
&&+\Et{-rMX+(\theta M+U)'(\beta+V)}\;\;\;-\Et{r\Gamma^{(3)}+(\theta M+U)'\beta}\;\;-r\Gamma^{(1)}_{s}X_{t}\\
&=&-r[p(s;t)-MX-\Gamma^{(2)}+\Et{MX+\Gamma^{(3)}}+\Gamma^{(1)}X_{t}]\\
&&-(\theta M+U)'V+\Et{(\theta M+U)'V}\\
&=&-r\bar p(s;t)-(\theta M+U)'V+\Et{(\theta M+U)'V}.
\end{eqnarray*}	
This proves (\ref{barp0}).

 \end{document}